\documentclass[a4paper,fleqn,review]{cas-sc}

\usepackage[numbers]{natbib}
\usepackage[title]{appendix}
\usepackage{xcolor}

\def\tsc#1{\csdef{#1}{\textsc{\lowercase{#1}}\xspace}}
\tsc{WGM}
\tsc{QE}

\usepackage{caption}
\usepackage{lineno}

\newcommand*{\dif}{\mathop{}\!\mathrm{d}}

\begin{document}
\let\WriteBookmarks\relax
\def\floatpagepagefraction{1}
\def\textpagefraction{.001}
\let\printorcid\relax
\shorttitle{PINN for Viscoelasticity}
\shortauthors{Z. Lin et~al.}

\title [mode = title]{A Physics-Informed Neural Network Framework for Simulating Creep Buckling in Growing Viscoelastic Biological Tissues}                      

\author[1]{{Zhongya Lin}}
\credit{Conceptualization, Methodology, Software, Analysis, Writing and reviewing manuscript}

\author[1]{{Jinshuai Bai}}
\cormark[1]
\ead{bjs@mail.tsinghua.edu.cn}
\credit{Conceptualization, Methodology, Software, Analysis, Writing and reviewing manuscript}

\author[1]{{Shuang Li}}
\credit{Software, Analysis, Writing and reviewing manuscript}

\author[1]{{Xindong Chen}}
\credit{Software, Writing and reviewing manuscript}

\author[1]{{Bo Li}}
\credit{Software, Analysis, Writing and reviewing manuscript}

\author[1,2]{{Xi-Qiao Feng}}
\cormark[1]
\ead{fengxq@tsinghua.edu.cn}
\credit{Supervision, Project administration, Funding acquisition, Writing and reviewing manuscript}

\affiliation[1]{organization={Institute of Biomechanics and Medical Engineering, AML Department of Engineering Mechanics, Tsinghua University},
	city={Beijing},
	citysep={}, 
	postcode={100084}, 
	country={China}}
\affiliation[2]{organization={Mechano-X Institute, Tsinghua University},
	city={Beijing},
	citysep={}, 
	postcode={100084}, 
	country={China}}

\cortext[cor1]{Corresponding author}

\begin{abstract}
Modeling viscoelastic behavior is crucial in engineering and biomechanics, where materials undergo time-dependent deformations, including stress relaxation, creep buckling and biological tissue development. Traditional numerical methods, like the finite element method, often require explicit meshing, artificial perturbations or embedding customised programs to capture these phenomena, adding computational complexity. In this study, we develop an energy-based physics-informed neural network (PINN) framework using an incremental approach to model viscoelastic creep, stress relaxation, buckling, and growth-induced morphogenesis. Physics consistency is ensured by training neural networks to minimize the system's potential energy functional, implicitly satisfying equilibrium and constitutive laws. We demonstrate that this framework can naturally capture creep buckling without pre-imposed imperfections, leveraging inherent training dynamics to trigger instabilities. Furthermore, we extend our framework to biological tissue growth and morphogenesis, predicting both uniform expansion and differential growth-induced buckling in cylindrical structures. Results show that the energy-based PINN effectively predicts viscoelastic instabilities, post-buckling evolution and tissue morphological evolution, offering a promising alternative to traditional methods. This study demonstrates that PINN can be a flexible robust tool for modeling complex, time-dependent material behavior, opening possible applications in structural engineering, soft materials, and tissue development.
\end{abstract}


\begin{highlights}
\item An energy-based PINN framework is developed for nonlinear viscoelasticity, encompassing creep buckling and tissue development.
\item Inherent training oscillations in PINN naturally trigger creep buckling of viscoelastic structures without artificial perturbations. 
\item Viscoelastic growth and morphogenesis simulations using energy-based PINN reveal complex folding patterns relevant to tissue development.
\end{highlights}


\maketitle

\section{Introduction}\label{sec1} 

\begin{figure}[htbp]
	\centering
	\includegraphics[width=0.9\textwidth]{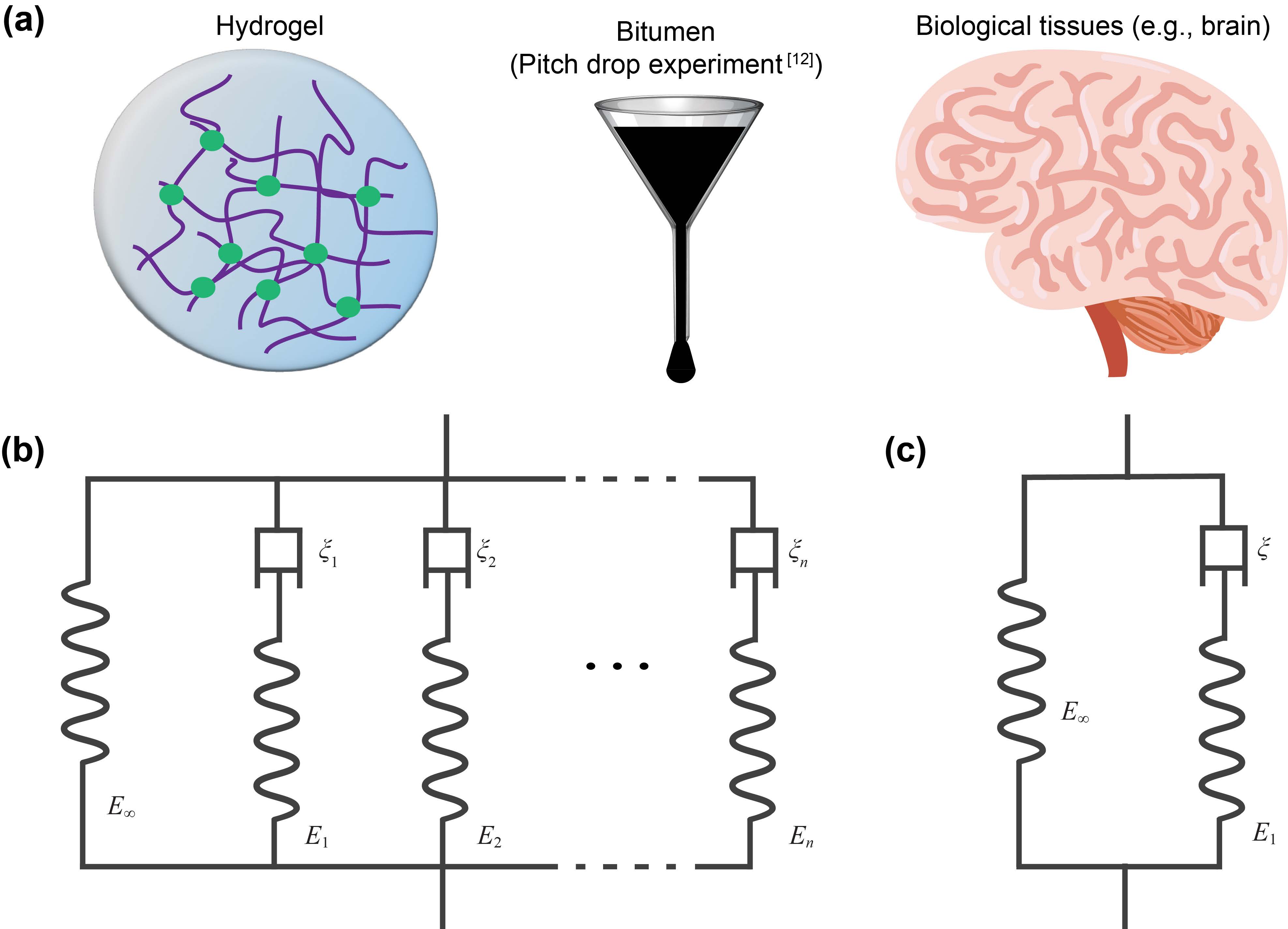}
	\caption{Viscoelastic materials and models. (a) Viscoelasticity is very common among different materials, e.g., hydrogels, bitumen \cite{Edgeworth.1984}, and soft biological tissues. (b-c) The viscoelastic models used for mechanical modeling. (b) The general Maxwell model. (c) The standard three-parameter model.}
	\label{fig:viscomodel}
\end{figure}

Many materials, including polymers, biological tissues and soft composites, exhibit combination of elastic energy storage and viscous energy dissipation \cite{Review_1}, i.e., the viscoelastic properties (Fig.\ref{fig:viscomodel}a). Understanding and predicting the time-dependent phenomena are essential for numerous engineering and biological applications \cite{Review_3}, such as designing advanced materials \cite{Lin.2020}, optimizing structural performance \cite{Sun.2013,Lin.2021}, and studying biological tissue growth and morphogenesis \cite{Collinet.2021}. The complex viscoelastic behaviors, particularly in cases involving large deformations, nonlinear responses, and evolving geometries, present persistent modeling challenges.

Over the past decades, computational approaches such as the finite element method (FEM) and finite difference method have been widely used to simulate viscoelastic responses, including creep, stress relaxation, buckling instabilities, and wave propagation \cite{Stewart.2024,Alves.2021,Reese.1998,Golla.1985}. While these methods have been successful in many scenarios, they often require refined meshing strategies and explicit perturbations for specific problems. For example, in the instability analysis, traditional FEM always need artificial perturbations or imperfections which can lead to different buckling modes \cite{Bazant.2010}. Moreover, in highly nonlinear and evolving systems, such as biological growth and morphogenesis, these conventional approaches can struggle to handle very complex geometries or require complex implementation strategies \cite{BenAmar.2025,Liu.2024a}. The need for a more adaptive computational framework is evident.

Physics-informed neural network (PINN) has recently emerged as a promising alternative to solve some numerical problems \cite{Raissi.2019, Review_2}. PINN offers a mesh-free approach that combines data and physical laws to solve differential equations by embedding the governing physics into the loss function of neural networks \cite{Raissi.2019, Wang.2024}. It has been successfully applied to forward and inverse problems in fluid dynamics \cite{Cai.2021, Zhu.2024}, solid mechanics \cite{Bai.2025, Zhuang.2021}, and material science \cite{Zhang.2022, Zhang.2024}, demonstrating their capability to solve physical problems. For the viscoelastic research, the inverse viscoelastic problem has been studied using a deep neural network to obtain parametric and neural-network-based viscoelastic models from limited displacement data \cite{Xu.2021}. A proposed ViscoelasticNet using PINN with appropriate constitutive models was applied to learn stress fields of viscoelastic fluids \cite{Thakur.2024}. The deep energy method was developed to solve partial differential equations governing the deformation of hyperelastic and viscoelastic materials \cite{Abueidda.2022}. It leverages the variational principles of mechanics, training networks to minimize a system's potential energy functional. These pioneering applications validate the use of PINN and related deep learning techniques for mechanical simulation and viscoelastic constitutive modeling. 

For instability studies, buckling analysis of Kirchhoff plates has been implemented using the deep energy method with the buckling governing equations-related loss function to predict the critical buckling load \cite{Zhuang.2021}. While this marks significant progress for elastic instabilities, the distinct phenomenon of creep buckling of viscoelastic structures—where instability develops over time under sustained load due to viscosity—necessitates a dedicated PINN formulation capable of handling long-term temporal integration and history-dependent material laws. Indeed, for viscoelastic structures, the creep buckling is one of the important failure modes, signifying failure after a period under steady load either via sudden snapping or excessive deflection \cite{Miyazaki.2015}. Additionally, during growth and development of soft biological tissues with viscoelastic property, the symmetry broken and geometrical stability loss are important processes for tissue morphogenesis \cite{Lin.2025, Moghe.2025, Harmansa.2023}. Despite the large potential of PINN, extending these approaches to capture creep buckling of viscoelastic structures, and to model the intricate interplay of mechanics and material evolution in viscoelastic growth and morphogenesis, presents difficulties and remains for investigation. 

Addressing these challenges, this study develops and applies an energy-based PINN framework for modeling viscoelastic behavior. We extend the energy-based PINN framework \cite{Bai.2025, Nguyen-Thanh.2020, Abueidda.2022, Zhuang.2021} to the setting of nonlinear viscoelasticity, where the governing equations depend not only on the instantaneous deformation but also on the full deformation history. This requires incorporating time-dependent evolution laws and coupling elastic and viscous energy contributions directly within the PINN formulation. Building on this foundation, we further examine how the dynamics of network optimization influence the numerical solution and show that optimizer-induced oscillations can serve as an intrinsic perturbation mechanism capable of initiating buckling in viscoelastic structures. In addition, we apply the framework to growth-induced deformation and morphogenesis, demonstrating that differential growth in cylindrical geometries can naturally lead to geometric instabilities and pattern formation. These results reveal that physics-informed neural networks can serve as an effective and flexible computational tool for simulating time-dependent material behavior and morphogenesis beyond the reach of conventional numerical methods. The remainder of this paper is organized as follows: Section 2 introduces the mathematical formulation of viscoelasticity. Section 3 details the implementation of PINN for viscoelastic simulation, and the benchmark of creep and stress relaxation. Section 4 explores the role of training oscillations in capturing creep buckling phenomena. Section 5 extends the framework to viscoelastic growth and morphogenesis, demonstrating its potential in biological applications. Finally, Section 6 discusses the advantages, limitations, and future research directions.

\section{Basic Equations for Viscoelastic Problems}

\subsection{Governing Equation of Nonlinear Viscoelasticity}

Let $\Omega$ be the configuration of a body in the three-dimensional Euclidean space $\mathbb{E}3$, and the corresponding initial configuration is $\Omega_0$. The deformation gradient tensor is $\mathbf{F}$, and $\mathbf{F}= \Delta \mathbf{u} + \mathbf{I}$, where $\mathbf{u}$ is the displacement vector and $\mathbf{I}$ is the identity tensor. For hyperelastic neo-Hookean materials, the stain energy density can be expressed as \cite{Ogden.1997}
\begin{equation}\label{eq:elastic_e}
	w = \frac G 2 \left(I_1 - 3 - 2 \ln J \right) + \frac \lambda 2 \left(J-1\right)^2,
\end{equation}
where $\lambda$ is the Lamé's first parameter, $G$ is shear modulus, $J= \text{det} \left( \mathbf{F} \right)$ is the Jacobian which measure the relative volume change, $I_1 = B_{ii}$ is the first deformation invariant. $\mathbf{B} = \mathbf{F} \cdot \mathbf{F}^T$ is the left Cauchy-Green deformation tensor. The Cauchy stress formula can be deduced by differentiating the strain gradient density, as 
\begin{equation}\label{eq:elastic_sigma}
	\sigma_{ij} = \frac 1 J F_{ik} \frac{\partial w}{\partial F_{jk}} = \frac{G}{J} \left( \mathbf{B} - \mathbf{I}\right) + \lambda \left( J -1 \right) \mathbf{I},
\end{equation}
where the relations $\partial I_1 / \partial F_{ij} = 2 F_{ij}$ and $\partial J / \partial F_{ij} = J F_{ij}^{-1}$ are used. 

For large-deformation viscoelastic model, based on previous studies \cite{Khajehsaeid.2014, Gamonpilas.2012}, the constitutive equation can be written as 
\begin{equation}\label{eq:visco_sigma}
	\sigma (t) = \int_{-\infty}^{t} m(t - \zeta) \frac{\partial}{\partial \zeta} \left(\frac 1 J \mathbf{F} \cdot \frac{\partial w}{\partial \mathbf{F}} \right) \dif \zeta,
\end{equation}
where $m(t)$ is the dimensionless relaxation function and not changes during deformation. Using the general Maxwell model (Fig. \ref{fig:viscomodel}b), $m(t)$ can be expressed by Prony series: 
\begin{equation}
	m(t) = m_{\infty} + \sum_i m_i \exp \left( -\frac {t}{\tau_i}\right),
\end{equation}
where $\tau_i=\xi_i / E_i$ are relaxation times, $m_{\infty} + \sum_i m_i = 1$. If we describe viscoelasticity by standard solid model (Fig. \ref{fig:viscomodel}c), it has
\begin{equation}
	m(t) = m_{\infty} + m_1 \exp \left( -\frac {t}{\tau}\right),
\end{equation}
where $m_{\infty} = E_{\infty} / \left(E_{\infty} + E_1\right)$, $m_1 = E_1 / \left(E_{\infty} + E_1\right)$, and $\tau = E_1 / \xi$.

If following the description of neo-Hookean hyperelasticity (Eq. (\ref{eq:elastic_sigma})) and the general viscoelastic form (Eq. (\ref{eq:visco_sigma})), the viscoelastic constitutive equation can be written as
\begin{equation}\label{eq:sigma}
	\sigma (t) = \int_{-\infty}^{t} G(t - \zeta) \frac{\partial}{\partial \zeta} \left(\frac{\mathbf{B} - \mathbf{I}} {J} \right) \dif \zeta + \int_{-\infty}^{t} \lambda(t - \zeta) \frac{\partial}{\partial \zeta} \left[\left( J -1 \right) \mathbf{I} \right] \dif \zeta,
\end{equation}
where $G(t)$ and $\lambda(t)$ are relaxation functions similar to the Lamé constants, assuming deviatoric and volumetric responses relax differently with isotropic relaxation functions.

\subsection{Stress Relaxation and Creep}

For viscoelastic stress relaxation condition, the deformation keeps constant, denoted as $\mathbf{F}_0$. Then, the Cauchy stress evolution equation can be obtained from Eq. (\ref{eq:sigma}), i.e.,
\begin{equation}\label{eq:relax0}
	\sigma (t) = G(t) \left[ \frac{\mathbf{B}_0 - \mathbf{I}} {J_0} \right] + \lambda(t) \left( J_0 -1 \right),
\end{equation}
where $\mathbf{B}_0 = \mathbf{F}_0 \cdot \mathbf{F}_0^T$, $J_0 = \det \left( \mathbf{F}_0 \right)$. Specially, under the small deformation and one-dimensional condition, the strain keeps constant as $\varepsilon_0$, and the stress evolution equation reduces to $\sigma(t) = E(t) \varepsilon_0$ \cite{Christensen.1982}. When using the standard solid model Fig. \ref{fig:viscomodel}c, the stress is 
\begin{equation}
	\sigma(t) = \left[ E_\infty + E_1 \exp\left(-\frac{t} {\tau} \right) \right] \varepsilon_0, 
\end{equation}
and the stress relation rate is 
\begin{equation}
	\dot{\sigma} (t) = - \frac{E_1 \varepsilon_0}{\tau} \exp\left(-\frac{t} {\tau} \right).
\end{equation}
The relaxation time $\tau$ means the needed time that the material relaxed all relatable stresses if the stress is relaxed with the initial rate $\dot{\sigma} (0)= - (E_1 \varepsilon_0)/{\tau}$. 

Creep is defined as the tendency of a solid material to deform permanently under the influence of constant stress. For creep condition, the stress keeps constant, i.e., $\sigma_0$. The strain creep function can be written as $\mathbf{e} (t) = J(t) \sigma_0$, where $J(t)$ is the creep compliance function and $\mathbf{e}$ is the strain tensor. The creep compliance can be obtained by applying the Laplace transform and inverting the relaxation function. It may be not solved analytically under large deformation, but the creep behavior can be captured easily by numerical method, like finite element analysis. Specially, for the small deformed one-dimensional case, the creep process becomes using the standard solid model (Fig. \ref{fig:viscomodel}c)
\begin{equation}\label{eq:creep1}
	\varepsilon(t) = \sigma_0 \left[ \frac{1}{E_\infty} - \frac{E_1}{E_\infty (E_\infty + E_1)} \exp \left( - \frac{t}{\tau_r}\right) \right],
\end{equation}
where the retardation time $\tau_r$ is defined as $\tau_r = \tau (E_\infty + E_1)/E_\infty$, which can characterized the creep behavior. The creep rate is 
\begin{equation}
	\dot{\varepsilon}(t) = \frac{\sigma_0 E_1}{\tau_r E_\infty (E_\infty + E_1)} \exp \left( - \frac{t}{\tau_r}\right),
\end{equation}
If the material creep with the initial creep rate $\dot{\varepsilon} (0) = {\sigma_0 E_1}/[{\tau_r E_\infty (E_\infty + E_1)}]$, the duration of the creep process from $\varepsilon(0) = \sigma_0 / (E_1 + E_\infty)$ to $\varepsilon(\infty) = \sigma_0 / E_\infty$ is $\tau_r$.

\subsection{Creep Buckling of Viscoelastic Structures}

Creep buckling refers to a specific type of structural failure that occurs in materials subjected to a sustained compressive load over an extended period \cite{Miyazaki.2015,Liu.2022}. As the material creeps, its effective rigidity decreases. This reduction can lower the critical buckling load, i.e., the maximum load that a structure can withstand before it buckles, leading to premature failure even when loads are lower than the instantaneous buckling thresholds.

Bifurcation buckling may occur for slender structures under axial compression. For a linear elastic slender cantilever, the critical buckling load can be easily obtained as 
\begin{equation}\label{eq:buckle-beam}
	P_\text{cb} = \frac{\pi^2 E I_g}{4 l^2},
\end{equation} 
where $l$ is the length of cantilever, $E$ is the Young’s modulus of the material and $I_g$ is the area moment of inertia. For the cross-section with area $A$, the critical pressure is 
\begin{equation}\label{eq:buckle-beam2}
	\sigma_\text{cb} = \frac{p_\text{cb}}{A} = \frac{\pi^2 E I_g}{4 A l^2}.
\end{equation}

For a linear elastic thin-walled cylindrical shell under uniform radial pressure, the critical buckling pressure can be estimated by \cite{Miyazaki.2015}
\begin{equation}\label{eq:buckle-shell}
	p_\text{cc} = \frac{2E}{1-\nu^2} \left(\frac{a}{d}\right)^3,
\end{equation}
where $d$ is the cylinder diameter, and $a$ is the wall thickness. If a structure, such as a slender beam or a thin-walled cylindrical shell, has viscoelastic properties, its stiffness will become smaller and the deformation will become larger under the constant external load with time. Thus, its critical buckling force will be smaller than its original state. For creep buckling analysis, we can easily use the quasi-static approach, that is, buckling occurs at time $t_{cr}$ when the applied pressure $p$ equals the critical elastic buckling load calculated using the current relaxed modulus $E(t_{cr})$ and Eqs. (\ref{eq:buckle-beam2}) or (\ref{eq:buckle-shell}).

\subsection{Viscoelastic Tissue Growth}

\begin{figure}[htbp]
	\centering
	\includegraphics[width=0.65\textwidth]{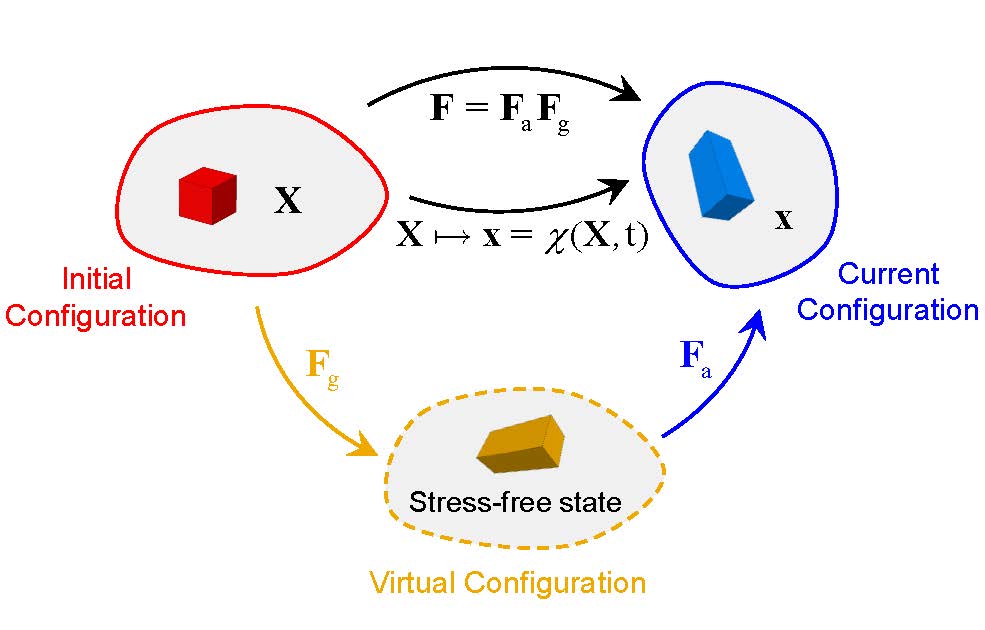}
	\caption{Diagram showing the decomposition of the deformation tensor when modeling tissue growth. The corresponding deformation gradient is decomposed into $\mathbf{F}_a$ and $\mathbf{F}_g$. A material point with the coordinate vector $\mathbf{X}$ in the initial configuration is mapped to $\mathbf{x}$ in the current configuration by $\chi(\mathbf{X}, t)$. The viscoelasticity leads to time-dependent deformation, which evolves with time.}
	\label{fig:growth-tensor}
\end{figure}

Viscoelasticity is a fundamental property of biological soft tissues, characterized by their ability to exhibit both elastic and viscous behavior when subjected to deformation \cite{Chaudhuri.2020, Clement.2017}. The stress modulated growth models have been developed \cite{Walker.2023, Sun.2022, Huang.2024} since the stress-growth law proposed by Fung \cite{Fung.1990}. In this kind of growth model, the deformation gradient tensor is decomposed as (Fig. \ref{fig:growth-tensor})
\begin{equation}\label{eq:F}
	\mathbf{F} = \mathbf{F}_a \cdot \mathbf{F}_g,
\end{equation}
where $\mathbf{F}_a$ is the mechanical deformation tensor and $\mathbf{F}_g$ is the growth tensor. The growth tensor is always written as $\mathbf{F}_g=g_i\mathbf{I}$, where $g_i$ is the growth ratio. The growth rate is determined by the stress state \cite{Huang.2024, Lin.2025}, i.e., 
\begin{equation}\label{eq:growth}
	\dot g_i = k \left(\sigma_i + b^g_{i} \right) g_i,
\end{equation}
where $\mathbf{b}^g = b^g_{i}\mathbf{I}$ is the biochemical forces driving growth \cite{Buskohl.2014}, and $k$ is a constant to regulate the stress effect on growth. For viscoelastic tissues, the stress $\sigma$ is obtained by Eq. (\ref{eq:sigma}) using deformation gradient tensor $\mathbf{F}_a$, which can relax as time goes. Stress relaxation slows residual stress accumulation, leading to increased growth rates \cite{Lin.2025}. 

A critical consequence of the differential or constrained growth described by the growth tensor $\mathbf{F}_g$, coupled with the viscoelastic response, is the generation of internal residual stresses \cite{Erlich.2024, Lin.2025}. When growth is non-uniform (e.g., faster surface growth relative to the bulk) or spatially constrained, significant compressive stresses can accumulate within the tissue over time. If these growth-induced compressive stresses reach a critical threshold, the initial configuration of the tissue may lose stability, leading to a buckling event \cite{Li.2011,Huang.2024}. This mechanical instability provides a fundamental mechanism for morphogenesis, driving the spontaneous formation of complex patterns such as wrinkles, folds, or creases observed during development and other biological processes.

Buckling analysis using the finite element method (FEM) often employs eigenvalue methods to determine the buckling load factor and associated mode shapes. In FEM, simulating the progression of instability, particularly time-dependent creep buckling, typically requires introducing perturbations or artificial imperfections. Additionally, modeling tissue growth within a finite element framework often requires the development of custom user programs capable of accurately simulating complex biological processes. In this study, we implement these problems using an energy-based physics-informed neural network framework, leveraging its potential to handle time-dependent properties and complex constitutive laws. 

\section{Numerical Implementation of Viscoelasticity via Energy-based PINN}

\begin{figure}[htbp]
	\centering
	\includegraphics[width=0.9\textwidth]{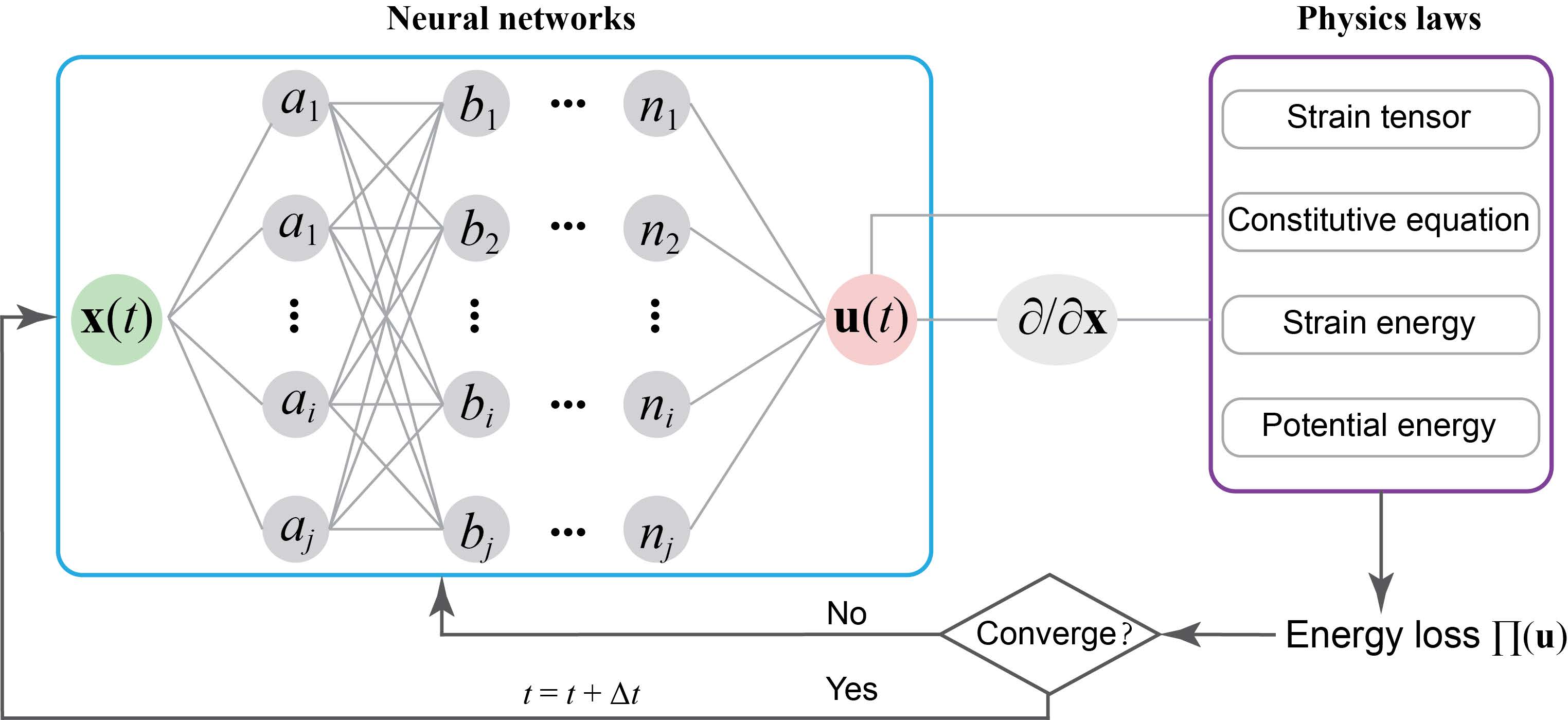}
	\caption{Diagram of the physics-informed neural networks.}
	\label{fig:newtork}
\end{figure}

This section details the numerical framework used to simulate the nonlinear viscoelastic phenomena described in the previous section. We employ an energy-based PINN approach, also known as the Deep Energy Method (DEM) \cite{Samaniego.2020, Abueidda.2022, Zhuang.2021, Bai.2025}. This method seeks the admissible displacement field that minimizes the total potential energy of the system. To handle the inherent time-dependency effect of viscoelasticity and growth, we utilize an incremental, time-stepping procedure, illustrated schematically in Fig. \ref{fig:newtork}.

\subsection{Energy-Based Formulation for Viscoelasticity}

In the energy-based PINN for viscoelasticity, the core idea is to approximate the displacement field $\mathbf{u}$ at discrete time points $t_i (i=0, 1, ..., n)$ using the neural networks. At each time step $t_i$, we solve a pseudo-elastic problem. The network takes spatial coordinates $\mathbf{x}$ as input and outputs the displacement vector at that specific time step:
\begin{equation}
	\mathbf{u} (\mathbf{x}, t_i) = N\left( \mathbf{x}; \mathbf{\varphi} \right),
\end{equation}
where $N(\cdot)$ is the neural network mapping with trainable parameters $\mathbf{\varphi}$ at time step $t_i$, and $\mathbf{x}$ is the position of material points. In practice, the neural network architecture, i.e. the number of hidden layers and the number of nodes per layer, is flexible and can be adjusted based on the PDE complexity \cite{Grossmann.2024, Nguyen-Thanh.2020, Li.2024, Bai.2023}. In this study, the neural network has three hidden layers and each layer contains twenty nodes. 

Before minimization at each time step $t_i$, the mechanical parameters $G(t)$ and $\lambda (t)$ are determined based on the elapsed time and the chosen viscoelastic model. For the three-parameter model used here (Fig. \ref{fig:viscomodel}c), there are $G(t_i) = G_{\infty} + G_1 \exp(-t_i/\tau)$ and $\lambda(t_i) = \lambda_{\infty} + \lambda_1 \exp(-t_i/\tau)$.  This operation treats the material as elastic within the time step but with stiffness properties that reflect the relaxation that has occurred up to time $t_i$. The potential energy of such pseudo-elastic system at time step $t_i$ can be formulated as  
\begin{equation}\label{eq:energy-func}
	\Pi(t_i) = \int_{V} \left[\frac {G(t_i)} {2} \left(I_1 - 3 - 2 \ln J \right) + \frac {\lambda(t_i)} {2} \left(J-1\right)^2\right] \dif V - \int_{V} \mathbf{f}(t_i)  \mathbf{u}(t_i) \dif V - \int_{S} \mathbf{p}(t_i) \mathbf{u}(t) \dif S,
\end{equation} 
where $\mathbf{f}$ is the body force, and the prescribed tractions $\mathbf{p}(t_i)$ on the boundary. The network parameters $\mathbf{\varphi}$ are found by minimizing the loss function, that is, the potential energy: 
\begin{equation}
	\mathbf{\varphi} = \arg \min_\mathbf{u} \mathcal{L} = \arg \min_\mathbf{u} \Pi(\mathbf{u}, G, \lambda, t_i),
\end{equation}
using the gradient descent method (e.g., Adam \cite{Kingma.2014}). Once converged, the displacement field $\mathbf{u}(\mathbf{x}, t_i)$ is obtained, and the corresponding stress tensor can be calculated using the constitutive equation. Then, the simulation advances to the next time step $t_{i+1}$ with increment $\Delta t$. This incremental process allows simulation of the full time-dependent viscoelastic response.

\subsection{Implementation of Boundary Conditions}

Traction boundary conditions are naturally incorporated through the external work term in the potential energy functional $\Pi$. For displacement (essential) boundary conditions, there are two main approaches to impose within energy-based PINN: the soft and hard ways. The soft way means adding a penalty term to the loss function that penalizes deviations from the prescribed displacement \cite{Barrett.1986, Bai.2025}. We use the hard enforcement to impose the displacement boundary conditions. This method imposes the boundary conditions exactly without adding penalty terms to the loss function, which can be beneficial for training convergence \cite{Bai.2025}. $\mathbf{u}(\mathbf{x}, t) = \mathbf{\bar u}(\mathbf{x}, t)$ for $\mathbf{x} \in \Gamma (\mathbf{x})$ can satisfy by modifying the displacement filed output by PINN through distance functions: 
\begin{equation}\label{eq:disp-bc}
	\mathbf{u}(\mathbf{x}, t_i) = N\left( \mathbf{x}; \mathbf{\varphi} \right) \odot b (\mathbf{x}) + \bar{\mathbf{u}} (\mathbf{x}, t_i),
\end{equation}
where $\odot$ means the element-wise product, and $b$ is an approximate distance function to the boundary $\Gamma (\mathbf{x})$ which denotes the shortest distance of a point $\mathbf{x}$ to the essential boundary. The distance function is non-negative and $b(\mathbf{x}) = 0$ only when $\mathbf{x} \in \Gamma (\mathbf{x})$. For the simple geometry with explicit boundaries, the method in Eq. (\ref{eq:disp-bc}) is easy to implement. For complex geometries, the boundary conditions can imposed using distance functions \cite{Sukumar.2022}. Different methods to impose displacement boundary conditions on PINN have been listed and compared in \cite{Berrone.2023}.

\subsection{Benchmark Test and Discussion}

Tensile creep and stress relaxation are fundamental benchmarks for studying viscoelasticity, providing a foundation for understanding time-dependent material responses under constant loading or deformation conditions. These phenomena are widely observed in engineering and biomechanics, playing critical roles in applications such as polymer mechanics, structural stability, and soft tissue mechanics. Due to their simplicity, analytical tractability, and practical relevance, we employ these problems to evaluate the performance of PINN in modeling viscoelastic behavior.

\begin{figure}[htbp]
	\centering
	\includegraphics[width=0.8\textwidth]{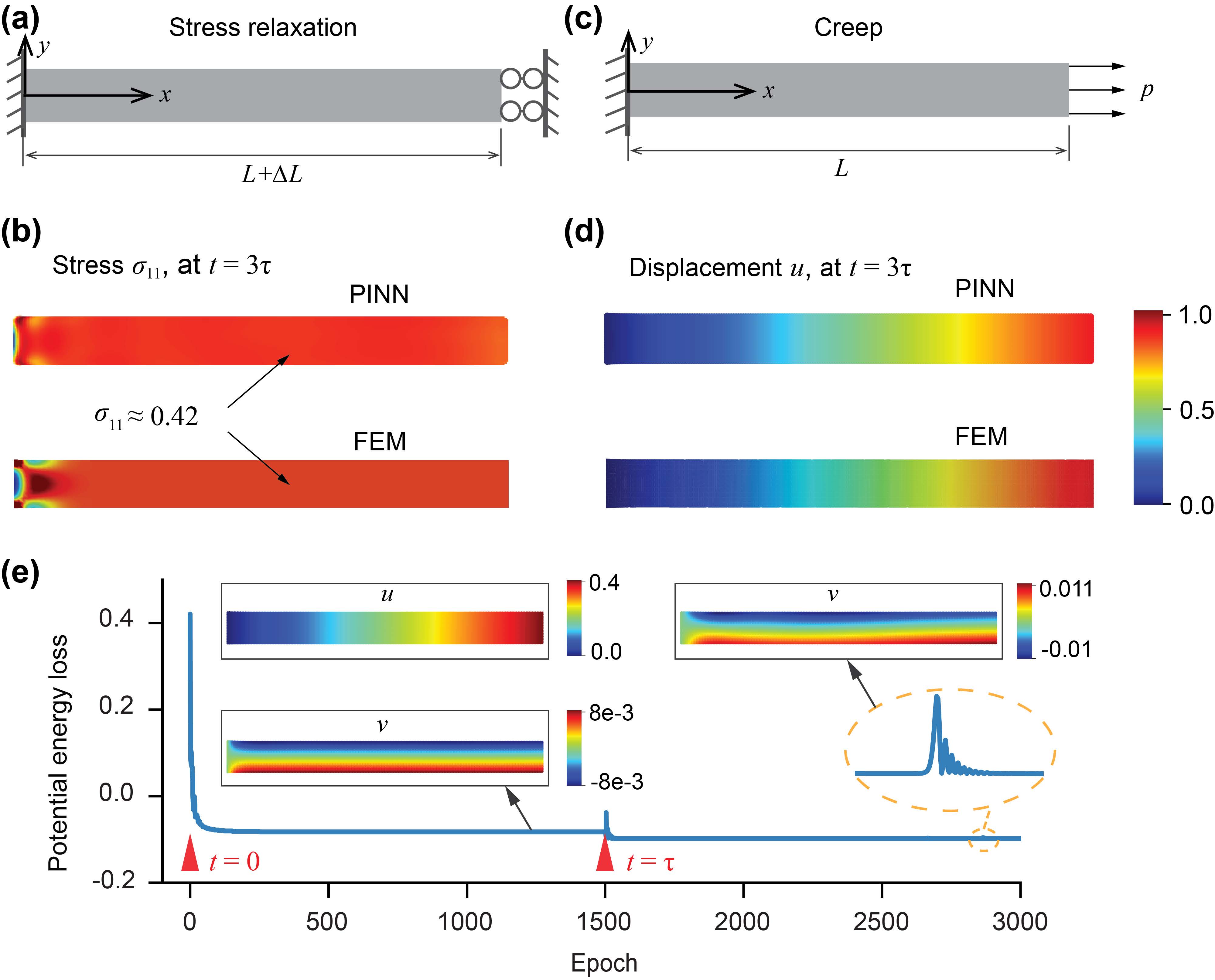}
	\caption{Viscoelastic creep and stress relaxation cases solved by PINN. (a) The schematic diagram of stress relaxation of a cantilever. The left boundary is fixed, and the right is fixed horizontally after stretching for a displacement $\Delta L$. (b) Comparison of the stress relaxation results at the time $t = 3 \tau$ from FEM and PINN. (c) The schematic diagram of creep of a cantilever. The left boundary is fixed, and the right boundary is subjected to a horizontal uniformly distributed tensile force. (d) Comparison of the creep results at the time $t = 3 \tau$ from FEM and PINN. (e) The training dynamics of the energy-based PINN. The potential energy loss decreases as training progress. Small oscillations may occur during the convergence process, which are brought in by the optimization algorithm itself.}
	\label{fig:benchmark}
\end{figure}

For stress relaxation, where a constant deformation is applied, the stress response is described by Eq. (\ref{eq:relax0}). We consider a cantilever beam (length $L = 10$ m, height $h=1$ m) subjected to a fixed tensile deformation $\Delta L = 1$ m imposed instantaneously at $t=0$, as illustrated in Fig. \ref{fig:benchmark}a. The moduli are $E_\infty = 4$ Pa, $E_1 = 6$ Pa, the viscosity parameter is $\xi=18$ Pa$\cdot$s, and the Poisson's ratio is $\nu=0.35$. Thus, the relaxation time is $\tau = \zeta / E_1 = 3$ s when using the standard linear viscoelastic model (Fig. \ref{fig:viscomodel}c). During the calculation, the relaxation functions still using the relations $G(t) = E(t)/2/(1+\nu)$, $\lambda(t) = \nu E(t)/(1+\nu)(1-2\nu)$, and the Poisson's ratio $\nu$ is time-independent. The fixed left end boundary and the right constant tensile deformation are implemented in the neural network, as
\begin{eqnarray}
	u_x(t) &=& N_x \left( \mathbf{x}, t; \mathbf{\varphi}_x \right) \cdot x \cdot (x-10) + x/10, \nonumber \\
	u_y(t) &=& N_y \left( \mathbf{x}, t; \mathbf{\varphi}_y \right) \cdot x.
\end{eqnarray}
The time increase is $\Delta t = 1.0$ s for each viscoelastic training. Fig. \ref{fig:benchmark}b compares the PINN-predicted stress distribution against FEM results, showing good agreement overall, with some deviations near the stress concentration at the fixed end.

The same beam geometry and material properties are used for tensile creep simulation (Fig. \ref{fig:benchmark}c). A constant tensile force corresponding to pressure $p=0.4$ Pa is applied to the right end ($x=L$) for $t \ge 0$. This traction boundary condition is included via the external potential which equals $p u _r h$. The displacement boundary is implemented in the neural network, as
\begin{eqnarray}\label{eq:bench-bcs}
	u_x(t) &=& N_x \left( \mathbf{x}, t; \mathbf{\varphi}_x \right) \cdot x, \nonumber \\
	u_y(t) &=& N_y \left( \mathbf{x}, t; \mathbf{\varphi}_y \right) \cdot x.
\end{eqnarray}
The viscoelastic creep displacement obtained from the PINN agree well with the finite element results (Fig. \ref{fig:benchmark}d). These benchmark results demonstrate that the implemented energy-based PINN framework, using an incremental time-stepping approach, accurately captures the fundamental time-dependent responses of viscoelastic materials.

\subsection{Oscillatory Convergence of Neural Network}

An interesting characteristic observed during the training of energy-based PINN is the oscillatory behavior in the convergence of the loss function or predicted displacements (Fig. \ref{fig:benchmark}e). These oscillations arise from the dynamics of the gradient descendant optimizer (e.g., SGD and Adam), which is very common during the deep learning. Such oscillations can facilitate escape from poor local minima but they may impede stable convergence during later optimization stages. In contrast, such oscillatory behavior may be advantageous for capturing phenomena like buckling in viscoelastic materials.

Classical numerical methods often require explicit artificial perturbations or initial imperfections to initiate buckling bifurcation from a perfect equilibrium path. We hypothesize that the inherent training oscillations within the energy-based PINN can act as a form of natural, internal perturbation. As the optimizer explores the energy landscape near a bifurcation point (where the landscape topology changes), these numerical oscillations can be sufficient to nudge the solution off the unstable primary path and guide it towards the physically stable, lower-energy buckled configuration. The potential applications are explored further in next sections. 

\section{Creep Buckling Analysis of Viscoelastic Structures via PINN}

This section presents numerical results for creep buckling of viscoelastic structures using the energy-based, incremental PINN. We demonstrate the effectiveness of this approach on two benchmark problems: (1) the creep buckling of a viscoelastic cantilever beam and (2) the creep buckling of a thin cylindrical shell under axial compression. A particular focus is placed on assessing the potential for PINN to capture instability-driven deformations by leveraging inherent training dynamics, potentially obviating the need for artificial perturbations often required in traditional methods. These examples highlight the capability of PINN to predict the time-dependent deformation and stability limits of viscoelastic structures.

\subsection{Creep Buckling of a Viscoelastic Cantilever Beam}

\begin{figure}[htbp]
	\centering
	\includegraphics[width=0.9\textwidth]{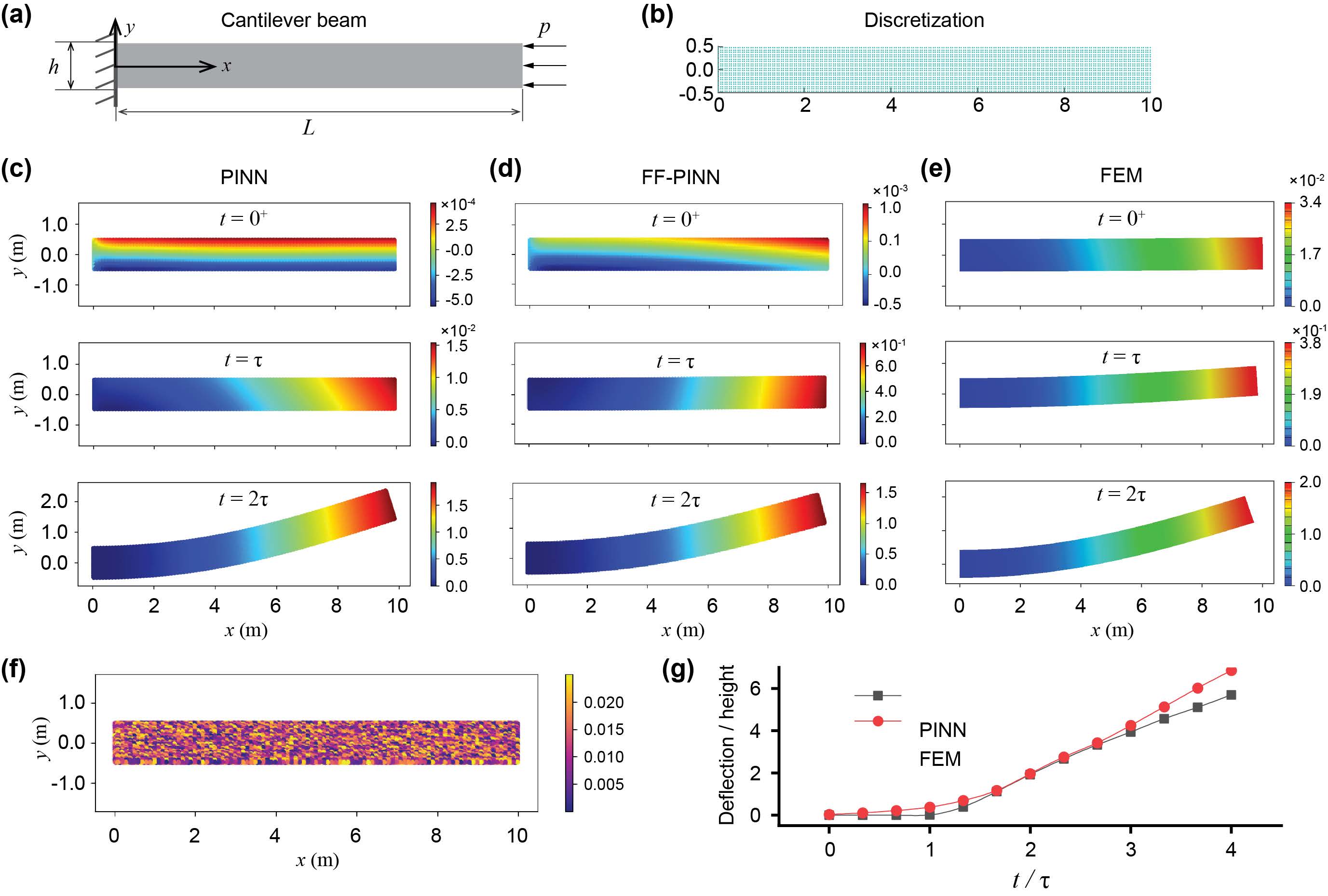}
	\caption{Creep buckling of a cantilever beam under axial compression. (a) Schematic diagram of a compressed cantilever beam. (b) Uniform discretization of the beam. (c) Displacement in the y-direction of the creep buckling cantilever beam at different times obtained from PINN under pressure $2\text{e}-2$ Pa, with Adam optimizer and learning rate $5\text{e}-3$. (d) Displacement in y-direction obtained from the Fourier feature-PINN (FF-PINN). The FF-PINN has three layers, and the feature number of each layer is 64 in this study. (e) Deformed cantilever beam obtained from FEM under pressure $2\text{e}-2$ Pa, with initial geometric imperfections. (f) Pointwise error of the y-displacement obtained from PINN and FEM at $t=2 \tau$. The error corresponds to the square error. (g) A comparison of the time-deflection curves obtained from PINN and FEM.}
	\label{fig:beam-buckling}
\end{figure}

Before analyzing the time-dependent creep buckling, we first validated the capability of our energy-based PINN framework to capture purely elastic buckling instabilities. We simulated the buckling of the cantilever beam geometry described in Fig. \ref{fig:beam-buckling}a, but using a hyperelastic constitutive model (described by Eqs. (\ref{eq:elastic_e}) and (\ref{eq:elastic_sigma})) reflecting the material's instantaneous elastic response under axial compression. The framework successfully predicted the Euler buckling mode shape, showing good agreement with finite element analysis (Fig. \ref{fig-a:beam-bucklemode}). Notably, the buckling bifurcation was captured naturally during the optimization process without requiring predefined geometric imperfections. This successful elastic benchmark provides confidence in the framework's ability to handle geometric nonlinearities relevant to buckling (Appendix A). Besides, as supporting evidence, simulations performed with a sub-critical load showed only minor oscillations during training without leading to buckling, as detailed in Appendix B.

Next, we consider the creep buckling behavior of a viscoelastic cantilever beam under uniaxial compression, as illustrated in Fig. \ref{fig:beam-buckling}a. The beam has height $h$=1.0 m and length $L$=10 m, with its left end clamped while the right end is subjected to a constant compressive load in the axial direction. The beam has the same mechanical parameters as the benchmark simultions of viscoelastic creep and stress relation, that is, the moduli are $E_\infty = 4$ Pa, $E_1 = 6$ Pa, the viscosity parameter is $\xi=18$ Pa$\cdot$s, and the Poisson's ratio is $\nu=0.35$. The pressure applied in the $x$-direction is set to $p = $2\text{e}-2 Pa (always kept in the $x$ direction), inducing time-dependent deformation and eventual buckling due to creep effects. Uniform collocation points discretize the domain, as shown in Fig. \ref{fig:beam-buckling}b. The $x$-direction displacement of the free end $u_r$ is obtained from the network, and the external potential equals $p u_r h$, which is included in the potential energy. The fixed left end boundary is implemented as Eq. \ref{eq:bench-bcs}. The simulation proceeds incrementally with time steps $\Delta t = 1$ s.

The evolution of the deformation over time is shown in Fig. \ref{fig:beam-buckling}c-e. As creep deformation progresses, the beam undergoes significant lateral displacement, leading to buckling. To assess the effect of Fourier feature embeddings on representational capacity, we computed the solution using a Fourier-Feature PINN (FF-PINN) \cite{WANG2021113938}, as shown in Fig. \ref{fig:beam-buckling}d. The FF-PINN produces deformation fields that are qualitatively similar to those of the baseline PINN and does not lead to a noticeable improvement for this problem. This is expected because the physical response is dominated by low-order global buckling modes, so the added high-frequency expressivity has limited influence on accuracy. The creep buckling with FEM is predicted by adding some initial geometric imperfections (Fig. \ref{fig:beam-buckling}e) under the same axial pressure. A quantitative comparison of the methods is provided in Fig. \ref{fig:beam-buckling}f, which shows the squared pointwise error fields between the PINN prediction and FEM solution at $t=2 \tau$. Fig. \ref{fig:beam-buckling}g compares the normalized mid-span deflection over time. The time–deflection curves of the PINN and FEM do not coincide at every time point once the structure enters the post-buckling regime. This discrepancy is expected: after the onset of buckling, the mechanical system becomes highly sensitive to small perturbations, and the post-buckling path is no longer unique. In FEM, the specific imperfection introduced to trigger buckling determines which post-buckling branch is followed. While in the PINN approach, the optimizer-induced perturbations naturally select a different, though mechanically admissible, buckling trajectory. Because these post-critical paths correspond to distinct but energetically similar equilibrium branches, perfect time-point agreement between PINN and FEM is not expected. Importantly, both methods predict the same buckling onset, overall mode shape, and qualitative evolution, which is a meaningful benchmark for creep-induced instability.

The PINN solution captures the gradual accumulation of strain and the onset of instability. To validate the predicted buckling behavior, we compare the eigenvalue of the viscoelastic cantilever with two reference cases obtained from FEM simulations (ABAQUS, plane strain analysis with CPE8R elements): one is modeled as an elastic material with initial modulus $E_0 = E_\infty + E_1$, representing the upper bound of stiffness, and another is modeled with its long-term modulus $E=E_\infty$, serving as a baseline reference. The eigenvalue of a beam with the infinity modulus ($E_{\infty}$) is obtained from finite element simulation, which is $9.36\text{e}-3$, and the beam with initial stiffness ($E_0$) is $2.34\text{e}-2$. The buckling pressure $2\text{e}-2$ for viscoelastic cantilever is between these two elastic conditions. These predictions confirm that the buckling pressure of the viscoelastic cantilever falls within the expected range between two elastic cases. Crucially, the buckling behavior in the PINN simulation emerged without introducing any explicit geometric imperfections or transient load perturbations. The transition to the buckled state appears to be triggered by the inherent dynamics of the optimizer exploring the solution space near the bifurcation point, consistent with the hypothesis outlined in Section 3.4. As supporting evidence, simulations performed with a sub-critical load ($p =5\text{e}-3 < p_\text{cr}$) showed only minor oscillations or shaking during training without leading to buckling, as detailed in Appendix B. This highlights a potential advantage of PINN for stability analysis.

\subsection{Creep Buckling of a Thin Cylindrical Shell under External Pressure}

\begin{figure}[!htbp]
	\centering
	\includegraphics[width=0.9\textwidth]{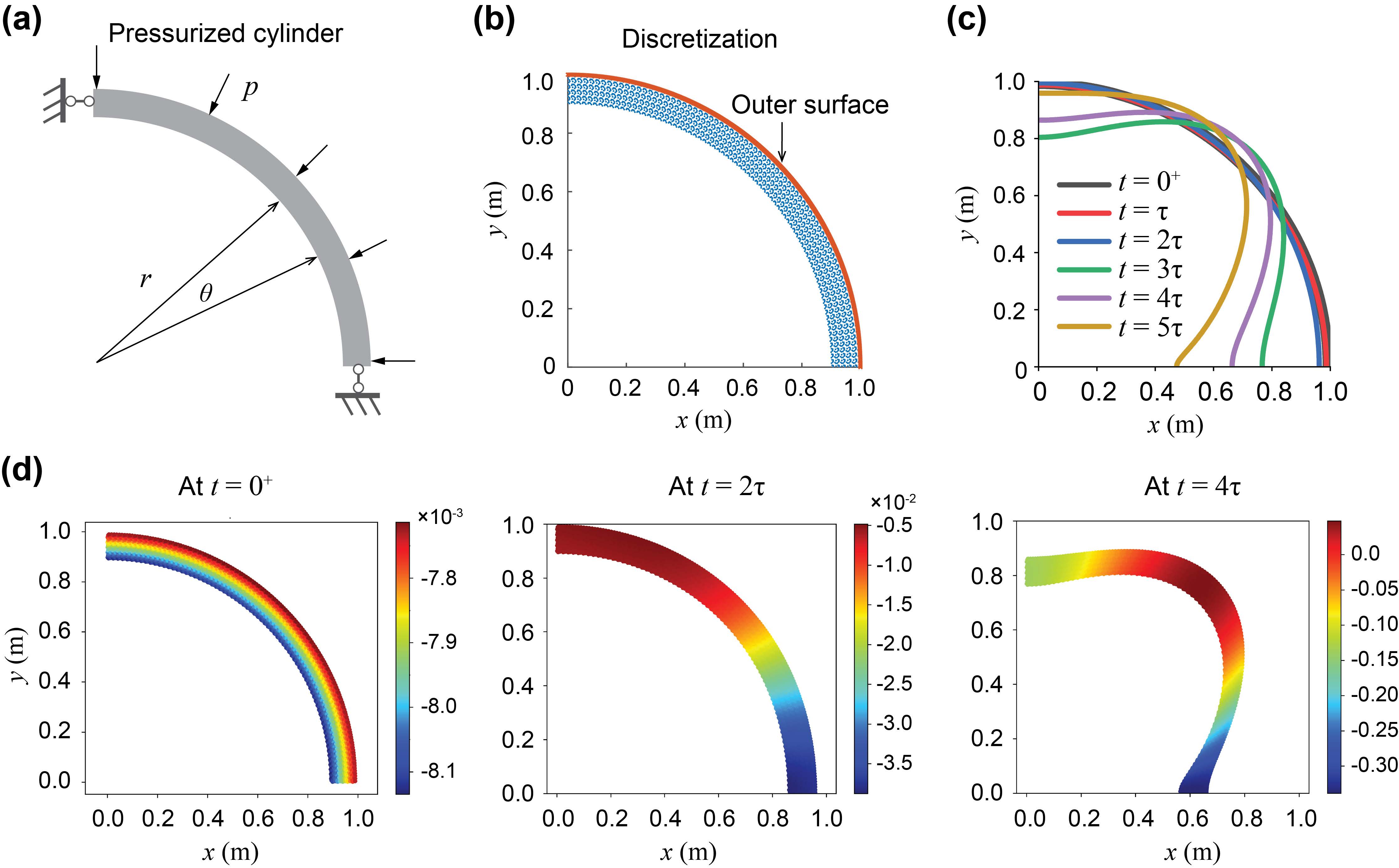}
	\caption{Creep buckling of a thin cylinder with outside radial pressure. (a) The schematic diagram of a compressed quarter cantilever cylinder, with the pressure equals to $1\text{e}-2$. (b) The discretization of the cylinder, and the integration at the outer surface. (c) The outer surfaces at different times. (c) Deformed quarter cylinder at different times. The forth order eigenvalue of a beam with the infinity modulus ($E_{\infty}$) is obtained from finite element simulation (Plain strain, CPE8R), which is $6.44\text{e}-3$, and the cylindrical shell with initial stiffness ($E_0$) is $1.61\text{e}-2$. The buckling pressure $1\text{e}-2$ for viscoelastic cantilever is between these two elastic conditions. }
	\label{fig:cylinder-buckling}
\end{figure}

Thin cylindrical shells under external pressure are susceptible to time-dependent instabilities due to viscoelastic creep. Due to symmetry, we model only a quarter of the cylindrical shell with appropriate symmetry boundary conditions applied on the cut surfaces ($\theta \in [0, \pi/2]$). The schematic representation of the problem is depicted in Fig. \ref{fig:cylinder-buckling}a, where an initially quarter cylindrical shell of outer radius $r_o=1$ m and thickness $a=0.1$ m is subjected to a uniform external pressure $p=1\text{e}-2$ Pa. The moduli are $E_\infty=4$ Pa, $E_1=6$ Pa, the viscosity parameter is $\xi=18$ Pa$\cdot$s, and the Poisson's ratio is $\nu=0.35$. 

The discretization of the shell domain is shown in Fig. \ref{fig:cylinder-buckling}b, where the uniform sampling points are used to resolve the spatial dependence of the displacement field. Sample points on the outer surface of the shell are used to calculate the outer surface displacement. The implement of this problem in PINN is under the cylindrical coordinate. The boundary condition is introduced to the neural network as
\begin{equation}
	u_{\mathbf{\theta}} = N\left( r, \theta; \mathbf{\varphi} \right) \cdot \sin \mathbf{\theta} \cdot \cos \mathbf{\theta}, 
\end{equation}
and the time increase is $\Delta t = 0.5$ s for each viscoelastic training.

The evolution of the shell's deformation is presented in Figs. \ref{fig:cylinder-buckling}c and \ref{fig:cylinder-buckling}d. In Fig. \ref{fig:cylinder-buckling}c, the deformation profile of the outer surface is plotted at different time instances, illustrating the progressive lateral deflection and the onset of buckling. The initially circular shell gradually deviates from its original shape due to viscoelastic creep, eventually leading to instability. The corresponding displacement fields at specific time snapshots are visualized in Fig. \ref{fig:cylinder-buckling}d. At $t=0^+$, the shell exhibits a nearly uniform radial contraction due to the applied pressure. As creep deformation progresses, localized bending deformations emerge, which intensify over time, culminating in a pronounced buckled shape at $t=4\tau$. To validate the PINN predictions, the buckling pressure obtained from the viscoelastic simulation is still compared with two classical elastic buckling solutions and FEM results. Case A: The shell is modeled as an material with initial modulus ($E=E_\infty + E_1$). Case B: The shell is modeled with its long-term elastic modulus ($E=E_\infty$). The comparison indicates that the buckling pressure of the viscoelastic cylinder within the expected range between two elastic cases.

The results presented for both the cantilever beam and the thin cylindrical shell confirm that the proposed PINN framework is a viable and accurate tool for predicting creep buckling phenomena of viscoelastic structures. The findings consistently suggest that PINN can capture these instability-driven deformations naturally through their intrinsic training dynamics, potentially streamlining stability analysis compared to traditional methods requiring explicit perturbations. Furthermore, we also proved theoretically how the Adam optimizer brings in perturabation for buckling (Appendix C). This capability highlights the significant potential of PINN for analyzing time-dependent instabilities in diverse engineering and biomechanics applications.

To evaluate the practical viability of the proposed method, we compare the wall-clock time of the PINN simulations with those of equivalent FEM simulations performed on the same workstation. Table 1 summarizes the computational cost for all benchmark problems. All problems are modelled on a CPU AMD Ryzen 7 5700G (3.80 GHz). Although the current PINN implementation is not computationally faster than FEM, it provides capabilities, such as mesh-free deformation, natural instability capture, and robust handling of evolving geometry, that are difficult for FEM to replicate. These complementary strengths justify PINN development despite the higher computational expense. Additionally, inference of the PINN is extremely fast, requiring only a forward evaluation of the neural network. The trained model can also be fine-tuned for nearby parameter regimes at a very low additional cost, amortizing the initial training effort across multiple related problems.

\begin{table}[t!]
	\centering
	\caption{A computational cost comparison of PINN and FEM for viscoelastic creep-buckling}
	\label{tb:timecomparison}
	\begin{tabular}{p{3.5cm} p{1.5cm} p{3cm} p{5cm}}
		\hline
		\textbf{Problem} & \textbf{Method} & \textbf{Mesh/Network size} & \textbf{CPU time for each time step} \\
		\hline
		{Cantilever beam buckling} & {FEM} &{1000 elements} &{< 1 s} \\
		{Cantilever beam buckling} & {PINN} &{$3 \times 20$ neurons} &{Train:$\simeq$ 3.8 mins; Inference: < 1 s} \\
		{Cylindrical shell buckling} & {FEM} &{1500 elements} &{< 1 s} \\
		{Cylindrical shell buckling} & {PINN} &{$3 \times 20$ neurons} &{Train:$\simeq$ 3.5 mins; Inference: < 1 s} \\
		\hline
	\end{tabular}
\end{table}

\section{PINN for Viscoelastic Growth and Buckling of Soft Biological Tissues}

Building upon the validated energy-based PINN framework, we now address the complex interplay between viscoelastic mechanics and growth in biological tissues, focusing on morphogenesis in cylindrical geometries. Structures like airways\cite{Li.2011, Wang.2023}, blood vessels, and intestines (Fig. \ref{fig:cylinder-growth}a), undergo growth driven by cellular proliferation and extracellular matrix remodeling. Understanding the mechanics governing these processes, particularly how growth patterns can lead to shape changes and instabilities, is vital for developmental biology, disease modeling, and tissue engineering. This section employs our energy-based PINN methodology to simulate growth in a viscoelastic cylindrical shell with outer radius $R_o=1$ cm and inner radius $R_i=0.9$ cm, and the outer surface is fixed, as depicted in Fig. \ref{fig:cylinder-growth}b. Due to assumed symmetry, we model half of the cylinder ($\theta \in [0, \pi]$) with symmetry boundary conditions on the cut surfaces (Fig. \ref{fig:cylinder-growth}b right). The material behavior is described by the viscoelastic constitutive model (specified in Section 2), with instantaneous short-term modulus $E_0 = E_\infty + E_1 =100$ kPa, long-term modulus $E_0=40$ kPa, the viscosity parameter $\xi=18$ kPa$\cdot$s, and the Poisson's ratio $\nu=0.49$.

In the implementation of energy-based PINN which minimizes the system's potential energy functional by approximating the displacement field, the total potential energy at time $t$ is written as 
\begin{equation}
	\Pi = \int_V \frac {G(t)} {2} \left(I^a_1 - 3 - 2 \ln J_a \right) + \frac {\lambda (t)} {2} \left(J_a-1\right)^2 \dif V + \int_V J_g b^g_i g_i \dif V,
\end{equation}
where the deformation gradient tensor is decomposed using Eq. (\ref{eq:F}),  $I^a_1 = \text{trace}(\mathbf{F}_a \cdot \mathbf{F}_a^T)$, $J_a = \det (\mathbf{F}_a)$, and $J_g = \mathbf{F}_g$. The solutions of displacement field $u(t)$, stress field $\sigma(t)$ and growth ratio $g_i(t)$ are found at discrete time points $t_k$. The process repeats for the next time step. To advance the growth ratio $g_i$ from time $t$ to $t + \delta t$, it is calculated using Eq. (\ref{eq:growth}), i.e.,
\begin{equation}
	g_i({t+\delta t}) = g_i(t) \exp[k(\sigma_i + b_i^g) \delta t],
\end{equation}
where $\delta t$ is the time increment for each iteration.

\subsection{Uniform Growth of a Cylindrical Tissue}

\begin{figure}[htbp]
	\centering
	\includegraphics[width=0.85\textwidth]{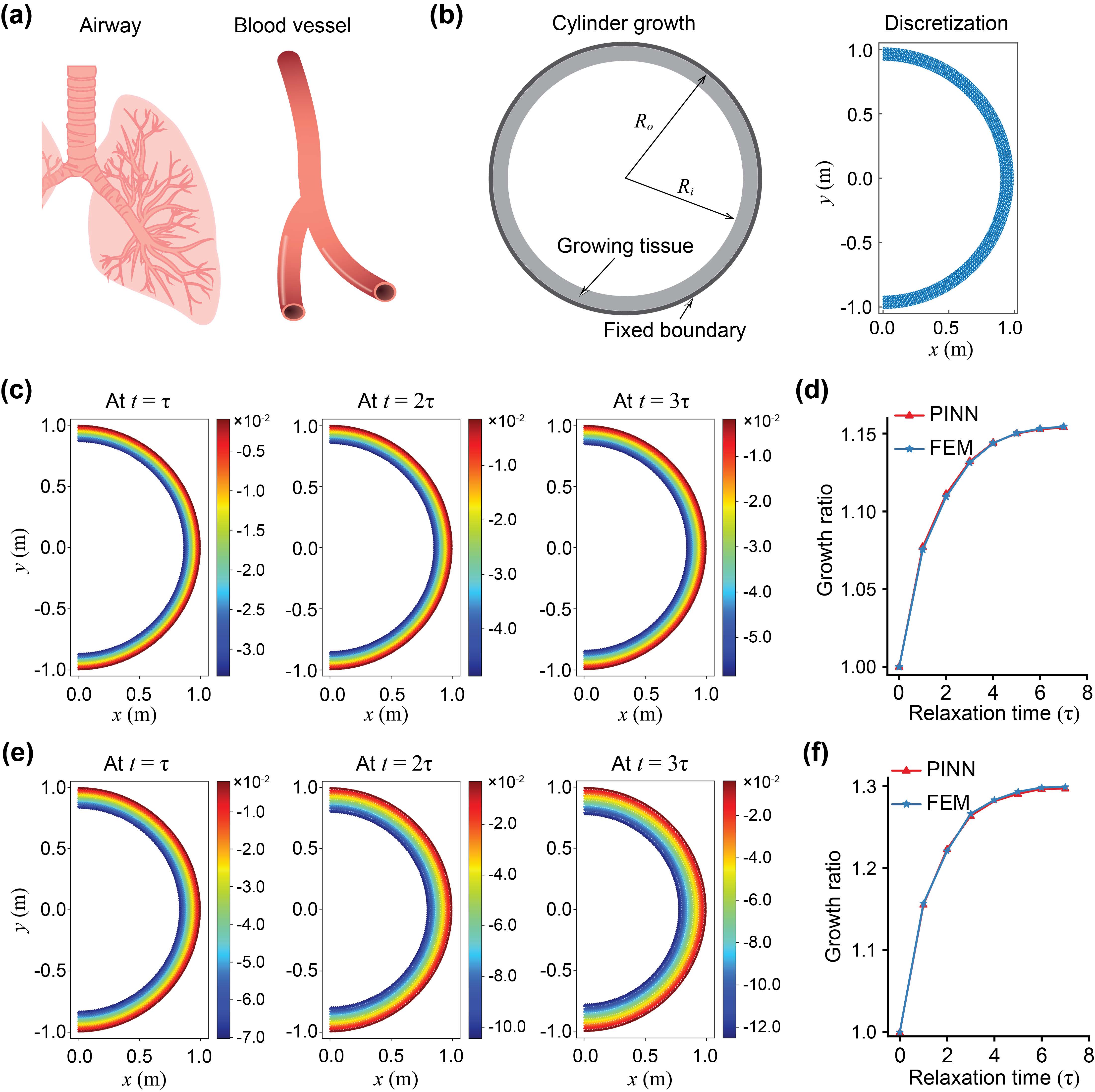}
	\caption{Cylinder symmetric growth with outer surface fixed. The growth process will proceed to the end, i.e., when the accumulated residual stress is equal to $b^g$. (a) Many biological tissues can be seen as a cylinder, e.g., the airway and blood vessel. (b) The schematic diagram and discretization of a semi-cylinder. (c) The total displacement at different times with biochemical stress $b^g=0.1 E_0$. (d) The comparison of growth ratio of the case (c) between PINN and FEM (ABAQUS with CPE4RH elements and user subroutine UMAT) results. (d) The comparison of growth ratio of the case (c). (e) The total displacement at different times with biochemical stress $b^g=0.2 E_0$. (f) The comparison of growth ratio of the case (e).}
	\label{fig:cylinder-growth}
\end{figure}

First, we consider that the tissue grows uniformly in both the radial and axial directions. The growth is assumed isotropic, implying the same growth rate in all directions, that is, the growth law is written as 
\begin{equation}
	\dot g = k (\sigma_{r} + \sigma_{\theta} + b^g) g,
\end{equation}
where the parameters are taken as $k=0.5 \, \text{s}^{-1} \cdot \text{kPa}^{-1}$, $b^g=10$ kPa. The boundary condition is introduced to the neural network as
\begin{align}
	u_{r} = N\left( r, \theta; \mathbf{\varphi} \right) \cdot (r- 1), \\
	u_{\theta} = N\left( r, \theta; \mathbf{\varphi} \right) \cdot \sin \mathbf{\theta} \cdot \cos \mathbf{\theta}.
\end{align}
Each viscoelastic training has 400 iterations, thus, the time increase is $\Delta t = 400 \delta t$ for each viscoelastic training. In this simulation, $\Delta t = 1$ s.

Figs. \ref{fig:cylinder-growth}c ($b^g = 0.1 E_0$) and \ref{fig:cylinder-growth}e ($b^g = 0.2 E_0$) depict the displacement magnitude at different times, where $\tau = 3$ s is the relaxation time. The color plots confirm that the cylindrical tissue expands uniformly in the radial and axial directions, reflecting isotropic growth. No signs of instability (wrinkling or buckling) are observed, as expected for uniform growth conditions. In Figs. \ref{fig:cylinder-growth}d and \ref{fig:cylinder-growth}f, we compare the growth ratio over time obtained from the PINN approach against FEM simulations. The results show excellent agreement between PINN and FEM, demonstrating the accuracy and robustness of the proposed approach. For tissues such as blood vessels or airways, uniform growth typically corresponds to healthy, symmetric development, free from pathological remodeling or localized aneurysmal expansions. PINN provide a mesh-free and flexible framework to study these scenarios without requiring heavy computational meshing and complex programming (e.g., UMAT for ABAQUS). Although no instability occurs under uniform growth, the development process may introduce localized deformations, potentially leading to wrinkling or bifurcation. This should consider differential growth at different directions, which can trigger complex pattern formations.

\subsection{Differential Growth, Buckling, and Morphogenesis}

In many biological systems, tissues undergo differential growth, wherein growth rates vary across different directions or tissue layers. This disparity often leads to mechanical instabilities that manifest as wrinkling, folding, or buckling. A well-known example is the morphogenesis of the chick foregut, which transitions from a relatively smooth cylindrical structure into a series of complex folds and luminal patterns over embryonic days (E8 to E17), as illustrated in Fig. \ref{fig:cylinder-buckling}a. These folded morphologies are critical for increasing surface area and optimizing nutrient transport and absorption. Consequently, understanding and predicting growth-induced buckling is essential for developmental biology, tissue engineering, and the design of bio-inspired soft materials.

For the differential growth, the tissue mechanical and geometrical parameters are the same as previous section. But the growth law is written as 
\begin{align}
	\dot g_r = k (\sigma_{r} + b^g_r) g_r, \\
	\dot g_{\theta} = k (\sigma_{\theta} + b^g_{\theta}) g_{\theta},
\end{align}
where the parameters are taken as $k=0.5 \, \text{s}^{-1} \cdot \text{kPa}^{-1}$, and $b^g_r = b^g_{\theta} = b^g$ in this simulation. This disparity in growth rates can lead to significant residual stresses and mechanical instabilities that shape the tissue morphology over time. During the simulation, the stress magnitude is set no greater than $b^g$.

\begin{figure}[htbp]
	\centering
	\includegraphics[width=0.95\textwidth]{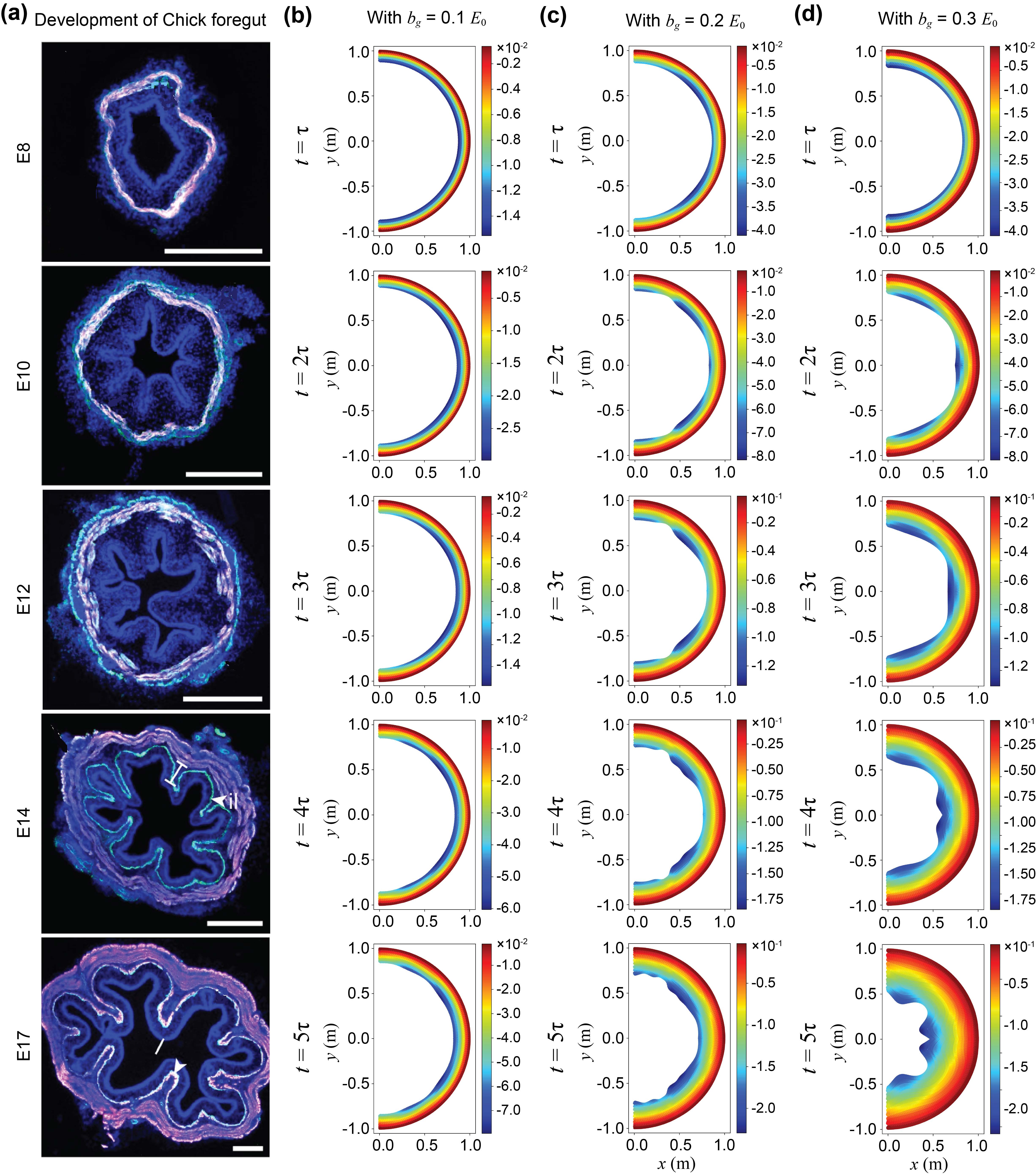}
	\caption{Cylinder growth with different growth ratios at the radial and circumferential directions. (a) Transverse sections over time in the chick foregut, adapted from \cite{Gill.2024}. (b) Growth morphology at different times with $b^g = 0.1 E_0$, (c) $b^g = 0.2 E_0$, and (d) $b^g = 0.3 E_0$.}
	\label{fig:growth-buckling}
\end{figure}

Figs. \ref{fig:cylinder-buckling}b-d illustrate the time evolution of the cross-sectional shapes. We vary the biochemical driving force ($b^g$) to represent different levels of differential expansion. For the mild differential growth (Fig. \ref{fig:cylinder-buckling}b), the cylinder remains largely axisymmetric at early times, but slight radial undulations emerge as growth progresses. The folds are shallow, indicating that while differential growth initiates instability, the magnitude of the growth mismatch is not sufficient to induce large-amplitude wrinkling. The moderate differential growth (Fig. \ref{fig:cylinder-buckling}c) shows more pronounced folding patterns. The radial expansion outpaces the circumferential expansion, creating residual hoop stresses that drive the tissue to buckle inward and outward periodically around the circumference. In Fig. \ref{fig:cylinder-buckling}d, the shell rapidly develops deep, well-defined folds at relatively early times ($t \simeq \tau$). During the differential growth, the number of folds may also increase. This bifurcation is common in tissue development \cite{Li.2011, BenAmar.2025}. Fig. \ref{fig:cylinder-buckling}a depicts cross-sectional slices of a chick foregut at different embryonic days. The progressive formation of circumferential folds resembles the morphologies observed in our PINN simulations under moderate to high differential growth. Although the exact geometry and material properties of embryonic tissues are far more complex, often involving multilayered structures and active cellular processes, the PINN-based model provides a mechanistic view of how differential growth can induce wrinkling and folding patterns reminiscent of in vivo development.

Consistent with the creep buckling results, the energy-based PINN framework captured the buckling instabilities leading to these patterns without requiring artificial perturbations. The optimizer navigating the energy landscape appears to naturally find the lower-energy buckled states when the initial configuration becomes unstable. These applications highlight the capability of energy-based PINN to handle complex coupled physics, including growth and viscoelasticity, and to naturally capture instabilities during the energy minimization process. This underscores their potential as a powerful computational tool for biomechanics and developmental biology.

\section{Discussion}

In this study, we developed and applied an energy-based PINN framework, utilizing the principle of minimum potential energy, to simulate complex viscoelastic phenomena. These included creep, stress relaxation, viscoelastic creep buckling, and growth-induced morphogenesis in cylindrical structures. By training neural networks to find displacement fields that minimize the system's potential energy functional within an incremental time-stepping scheme, our approach offers a mesh-free and flexible alternative to conventional numerical methods for these time-dependent problems. Furthermore, as demonstrated, it allows for the capture of buckling instabilities without the manual introduction of perturbations often required in finite element analyses.

A key highlight of our work is the natural emergence of buckling instabilities in creep simulations. Unlike traditional methods that typically rely on artificially imposed imperfections to trigger buckling, the inherent training dynamics of the neural network optimizer exploring the energy landscape appear to serve as a built-in mechanism for capturing the transition from stable creep to buckled configurations. This characteristic is particularly advantageous when modeling viscoelastic structures, where the gradual accumulation of strain leads to complex instability patterns. Furthermore, the extension of our framework to simulate tissue growth and morphogenesis underscores its versatility. By incorporating differential growth rates, the model successfully replicates the formation of intricate folding patterns reminiscent of those observed in biological tissues, thereby providing valuable insights into the mechanics of morphogenesis and offering potential applications in tissue engineering and bio-inspired design. In this study, we also using Fourier feature embeddings into the PINN. The results show that the FF-PINN does not significantly improve the accuracy or convergence for the viscoelastic creep-buckling problems considered in this work. We think the main reason is that the dominant physical behavior involves low-order global buckling modes mainly, which do not require high-frequency representational capacity.

Despite these promising advances, the study also reveals several limitations that warrant further investigation. For instance, while the framework effectively captures a buckling mode through simulation, it is currently not equipped to conduct a comprehensive eigenvalue buckling analysis as performed by traditional finite element methods, which can identify multiple potential buckling modes and their associated critical conditions. The natural perturbation is a strong, evidence-based hypothesis that needs further theoretical investigation into the link between optimizer dynamics and bifurcation theory. Additionally, the iterative nature of the incremental viscoelastic training loop, particularly when fine time-step resolution is required, may lead to significant computational cost compared to highly optimized implicit FEM solvers. These suggest that further optimization of the training process—possibly through adaptive time-stepping strategies or enhanced neural network architectures—will be necessary to fully exploit the potential of PINN in highly nonlinear or multiscale scenarios. Moreover, operator learning would amortize the computational cost of many-query tasks (design optimization, uncertainty quantification, digital twins), but applying it to nonlinear, history-dependent viscoelasticity introduces new challenges: the input space must represent the past deformation history or internal variables, training requires large, representative ensembles of solutions across parameter and history space, and careful incorporation of energy/thermodynamic constraints is necessary to preserve physical consistency. In future work, we can investigate strategies to represent memory in operator architectures, to reduce data requirements via transfer learning, and to combine energy-based losses with operator training to retain stability and thermodynamic consistency.

In summary, our study demonstrates that energy-based physics-informed neural networks hold significant promise for advancing the simulation of viscoelastic phenomena and growth-induced instabilities. The ability to naturally capture buckling behavior without external perturbations, combined with the flexibility to model complex tissue growth patterns within an energy-minimization framework, represents a noteworthy innovation in computational mechanics. However, addressing current limitations related to comprehensive stability analysis and computational efficiency will be essential for broadening the applicability of this approach in both research and practical engineering applications.

\section*{Declaration of competing interest}
The authors declare no competing interests.

\section*{Acknowledgments}
Support from the National Natural Science Foundation of China (Grant nos. 12032014, T2488101 and 12502234 ) is acknowledged.

\printcredits
\begin{appendices}
\renewcommand\thefigure{S\arabic{figure}}
\setcounter{figure}{0}  

\section{Validation Benchmark: Buckling Mode of a Hyperelastic Cantilever Beam}
To verify the fundamental capability of the implemented energy-based PINN framework to capture structural buckling, we simulated the classic Euler buckling of a hyperelastic cantilever beam under axial compression. The cantilever beam geometry (length $L=10$ m, height $h= 1$ m, assuming plane strain) and boundary conditions (clamped at $x=0$, axial pressure applied at $x=L$) were identical to those described in Section 4.1 of the main text. The material was modeled as compressible Neo-Hookean hyperelastic (described by Eqs. (\ref{eq:elastic_e}) and (\ref{eq:elastic_sigma})), using the instantaneous elastic properties of the viscoelastic material: Young's Modulus $E= E_1 + E_{\infty} = 10$ Pa and Poisson's ratio $\nu = 0.35$. The theoretical Euler critical buckling pressure for this configuration can be calculated by Eq. (\ref{eq:buckle-beam2}) , which is $2.06\text{e}-2$ Pa. The eigenvalue analysis of FEM (with ABAQUS, Buckle step, element CPE8R) is $2.34\text{e}-2$ Pa for first buckling mode and $0.203$ Pa for second buckling mode. 
\begin{figure}[htbp]
	\centering
	\includegraphics[width=0.85\textwidth]{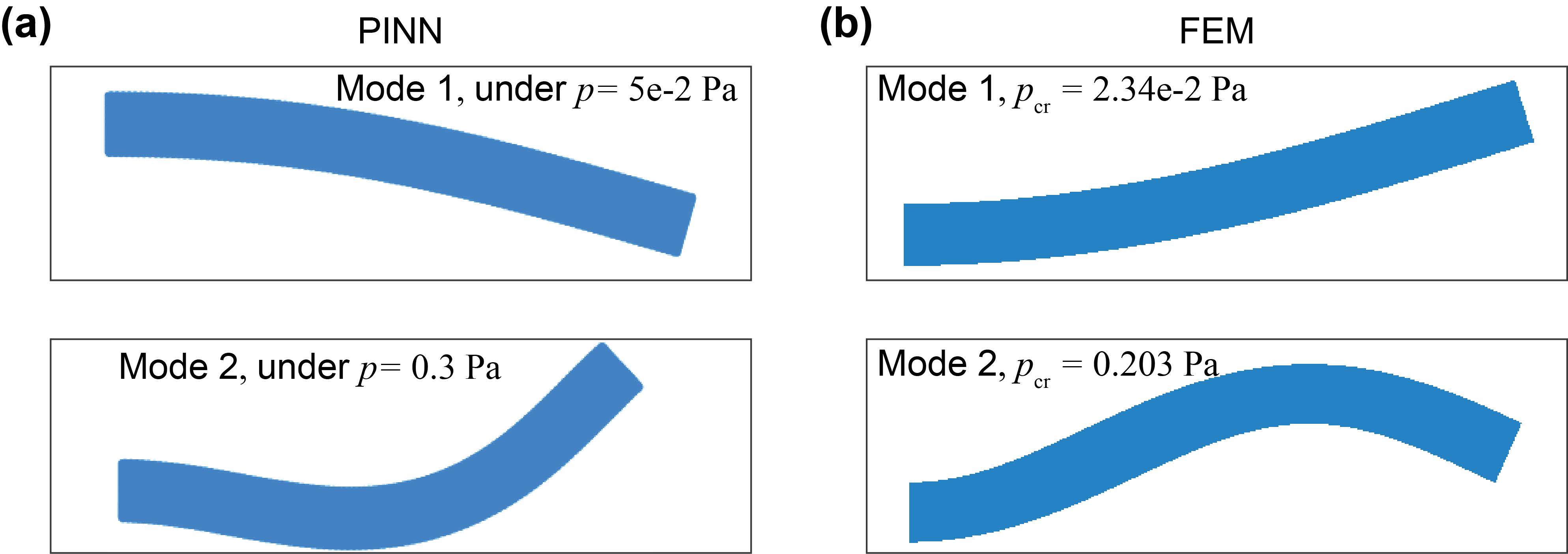}
	\caption{The buckling modes predicted by (a) PINN and (b) FEM.}
	\label{fig-a:beam-bucklemode}
\end{figure}

Fig. \ref{fig-a:beam-bucklemode} visualizes the buckling mode shape predicted by the PINN just after the bifurcation point, which matches the expected first and second Euler buckling modes for a cantilever beam. Significantly, this buckling mode was captured without introducing any artificial geometric imperfections; the numerical process of finding the minimum energy state successfully navigated the bifurcation. However, capturing higher-order buckling modes was not achieved with the current setup and requires further investigation. 

\section{Simulation under Sub-Critical Pressure}
To further investigate the role of neural network training dynamics in capturing buckling phenomena, we performed an additional simulation of the hyperelastic cantilever beam under a load known to be below the critical buckling threshold for the simulated time frame. The geometry, material properties, boundary conditions, and the energy-based incremental PINN framework were identical to those used for the buckling analysis in Appendix A. The applied compressive pressure is set to a sub-critical value of $p = 1\text{e}-2$ Pa, which is smaller than the critical buckling pressure ($p_{cr} = 2.34\text{e}-2$ Pa). 

\begin{figure}[htbp]
	\centering
	\includegraphics[width=0.9\textwidth]{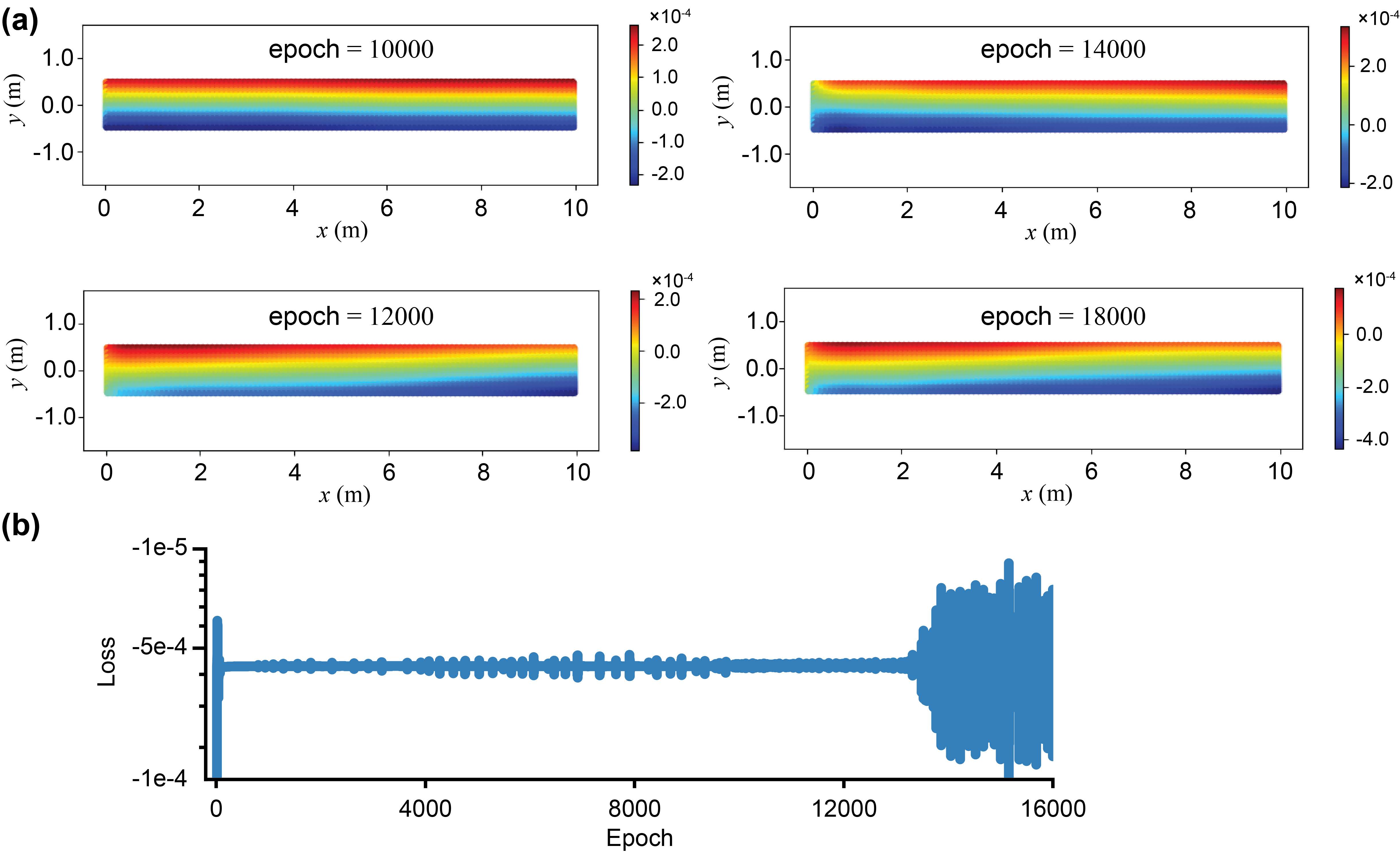}
	\caption{The deformed cantilever beam when the compressive pressure is lower than the buckling threshold. (a) The displacement in the y-direction of the deformed beam. (b) The convergence plot of training process.}
	\label{fig-a:beam-oscillation}
\end{figure}

Fig. \ref{fig-a:beam-oscillation}a shows the predicted displacement at the $y-$direction of the cantilever beam. Unlike the simulation under the critical load (Fig. \ref{fig:beam-buckling}c), there is no sign of rapid divergence or acceleration in deflection characteristic of buckling instability. The beam only develops some oscillatory deformation under perturbation. Fig. \ref{fig-a:beam-oscillation}b shows a representative convergence plot (the loss vs. training iterations). Oscillations during the optimization process are still evident, similar in nature to those observed during the buckling simulations.

The results demonstrate that even though training oscillations are present during the optimization process (as shown in Fig. \ref{fig-a:beam-oscillation}b), they do not lead to buckling instability when the applied load is below the critical threshold. This provides strong supporting evidence for the hypothesis presented in Sections 3.4 and 4.1: the training dynamics act as physically relevant perturbations that can trigger buckling only when the system is near its stability limit. They do not appear to be arbitrary numerical noise that destabilizes an otherwise stable configuration.

\section{Natural dynamics of Adam optimizer}

The Adam optimizer combines the momentum algorithm and adaptive scaling scheme during training \cite{kingma2014adam}. It updates the parameters as follows:
\begin{equation}
    \theta_{t+1}= \theta_{t} - \frac{\alpha}{\sqrt{\hat{v_t}}+\epsilon}\cdot \hat{m_t},
    \label{AEq1}
\end{equation}
\noindent where $\theta$ represents the trainable parameters in neural networks, $\alpha$ represents the learning rate, $\epsilon$ is a numerical stabilizer that prevents division by zero and is normally of $1\times10^{-8}$, $\hat{m_t}$ and $\hat{v_t}$ are the first and second momentum, respectively. The two momentum factors are updated by
\begin{equation}
\begin{aligned}
    \hat{m_t} &= \frac{m_t}{1-\beta_1^t}, \\ 
    \hat{v_t} &= \frac{v_t}{1-\beta_2^t}, \\
\end{aligned}
\label{AEq2}
\end{equation}
\noindent where
\begin{equation}
\begin{aligned}
    m_{t+1} &= \beta_1\cdot m_{t}+(1-\beta_1)\cdot g_t, \\
    v_{t+1} &= \beta_2\cdot v_{t}+(1-\beta_2)\cdot g_t^2.
\end{aligned}
\label{AEq3}
\end{equation}
In Eq.\ref{AEq3}, $\beta_1$ and $\beta_2$ are two parameters in the Adam optimisers and are defaultly set as 0.9 and 0.999, $g_t$ is the gradient of loss function, e.g., the overall energy of the system 
\begin{equation}
    g_t=\nabla_{\theta} \Pi.
    \label{AEq4}
\end{equation}
It is clear to find that the correction of the two momenta is effective at the beginning of training. Given that the training of PINN for viscoelastic problems require large number of training epochs, we simplify and ignore Eq.\ref{AEq2} in the following analysis.

Since $m_t$ and $v_t$ are all initialised as zero vectors, Eq.\ref{AEq3} can be also expressed as
\begin{equation}
    \begin{aligned}
        m_{t+1} &= (1-\beta_1) \sum_{n=0}^t{\beta_1}^n \cdot g_{t-n} \\
                &= (1-\beta_1) \sum_{n=0}^t{\beta_1}^n \cdot \nabla_{\theta} \Pi_{t-n}\\
        v_{t+1} &= (1-\beta_2) \sum_{n=0}^t{\beta_2}^n \cdot g_{t-n}^2 \\
                &= (1-\beta_2) \sum_{n=0}^t{\beta_2}^n \cdot (\nabla_{\theta} \Pi_{t-n})^2.
    \end{aligned}
    \label{AEq5}
\end{equation}
By substituting Eq.\ref{AEq5} into Eq.\ref{AEq1}, one can obtain
\begin{equation}
    \begin{aligned}
        \theta_{t+1}&=\theta_t-\frac{\alpha}{\sqrt{v_t}+\epsilon}\cdot m_{t+1}\\
        &=-\frac{\alpha}{\sqrt{v_t}+\epsilon}\cdot (1-\beta_1) \sum_{n=0}^t{\beta_1}^n \cdot (\nabla_{\theta} \Pi_{t-n}) \\
        &=-\frac{\alpha(1-\beta_1)}{\sqrt{v_t}+\epsilon}\cdot (\nabla_{\theta} \Pi_{t} + \sum_{n=1}^t{\beta_1}^n \cdot \nabla_{\theta} \Pi_{t-n})
    \end{aligned}
    \label{AEq6}
\end{equation}
Let $\tau_t=\frac{\sqrt{v_t}+\epsilon}{(1-\beta_1)}$ and $\Phi_t=-\sum_{n=1}^t{\beta_1}^n \cdot\nabla_{\theta} \Pi_{t-n}$, Eq.\ref{AEq6} can be simplifed and formulated as
\begin{equation}
    \theta_{t+1}-\theta_t=-\frac{\alpha}{\tau_t}\cdot(\nabla_{\theta} \Pi_{t}-\Phi_t).
    \label{AEq7}
\end{equation}
Eq.\ref{AEq7} is the stepwise parameter update rule. The continuous version of Eq.\ref{AEq7} (when $\alpha\to0$) \cite{wang2021understanding} is given as
\begin{equation}
    \begin{aligned}
        &\tau_t\cdot\dot\theta_t=-\nabla_{\theta} \Pi_{t}+\Phi_t,\\
        \Rightarrow &0=-\nabla_{\theta} \Pi_{t}-\tau_t\cdot\dot\theta_t+\Phi_t.
    \end{aligned}
    \label{AEq8}
\end{equation}
This equation is similar to the governing equation of the Langevin dynamics, which reads
\begin{equation}
    M\ddot{x}=-\nabla U(x)-\gamma\dot{x}+R_t, 
    \label{AEq9}
\end{equation}
where  $M$ is the mass of a particle in the Langevin dynamics, $U(x)$ denotes a potential, $R_t$ is a random white noise satisfying the Gaussian distribution. As observed, Eq.\ref{AEq8} lacks the inertia term compared to Eq.\ref{AEq9} and therefore is known as the \textit{underdamped} condition. During the training processes, the trainable parameters are updated by the Langevin dynamics, approching the minimal of energy potential. Meanwhile, the weighted mean of previous energy gradients, $\Phi_t$, acts as a perturbation during the training, preventing the trainable paramters trapped by saddle points. Moreover, $\tau$ is equivalent to the damper $\gamma$. From the dynamics point of view, this damper stabilises training and reduces loss oscillations. From the viewpoint of deep learning, $\tau$ is also an adaptive scaling vectors that balance the learning rate of different trainable parameters, assigning slowly-changing parameters a small damper and a large damper for frequently-changing parameters.

\end{appendices}

\bibliographystyle{cas-num}
\bibliography{cas-refs}

\end{document}